\documentclass[repr,
superscriptaddress,
 amsmath,amssymb,
 aps,
twocolumn,
 prl,
]{revtex4}
\usepackage[american]{babel}
\usepackage{graphicx}
\usepackage{times}
\usepackage{subfigure}
\usepackage{color}
\usepackage{hyperref}

\usepackage{mathpazo} % math & rm
\usepackage[scaled]{helvet} % ss
\usepackage{courier} % tt
\normalfont
\usepackage[T1]{fontenc}

\begin{document}
\title{Demonstrating various quantum effects with two entangled laser beams}

\author{B. Hage}
\affiliation{
ARC Centre of Excellence for Quantum-Atom Optics, Department of Quantum Science,The Australian National University, Canberra, Australian Capital Territory 0200, Australia}
\affiliation{ Centre for Quantum Computation and Communication Technology, Department of Quantum Science,The Australian National University, Canberra, Australian Capital Territory 0200, Australia}

\author{ J. Janou\v{s}ek}
\affiliation{
ARC Centre of Excellence for Quantum-Atom Optics, Department of Quantum Science,The Australian National University, Canberra, Australian Capital Territory 0200, Australia}
\affiliation{ Centre for Quantum Computation and Communication Technology, Department of Quantum Science,The Australian National University, Canberra, Australian Capital Territory 0200, Australia}

\author{ S. Armstrong}
\affiliation{
ARC Centre of Excellence for Quantum-Atom Optics, Department of Quantum Science,The Australian National University, Canberra, Australian Capital Territory 0200, Australia}
\affiliation{ Centre for Quantum Computation and Communication Technology, Department of Quantum Science,The Australian National University, Canberra, Australian Capital Territory 0200, Australia}

\author{ T. Symul}
\affiliation{ Centre for Quantum Computation and Communication Technology, Department of Quantum Science,The Australian National University, Canberra, Australian Capital Territory 0200, Australia}

\author{ J. Bernu }
\affiliation{ Centre for Quantum Computation and Communication Technology, Department of Quantum Science,The Australian National University, Canberra, Australian Capital Territory 0200, Australia}

\author{ H.M. Chrzanowski }
\affiliation{ Centre for Quantum Computation and Communication Technology, Department of Quantum Science,The Australian National University, Canberra, Australian Capital Territory 0200, Australia}
\author{ P.K. Lam }
\affiliation{
ARC Centre of Excellence for Quantum-Atom Optics, Department of Quantum Science,The Australian National University, Canberra, Australian Capital Territory 0200, Australia}
\affiliation{ Centre for Quantum Computation and Communication Technology, Department of Quantum Science,The Australian National University, Canberra, Australian Capital Territory 0200, Australia}
\author{ H.A. Bachor}
\affiliation{
ARC Centre of Excellence for Quantum-Atom Optics, Department of Quantum Science,The Australian National University, Canberra, Australian Capital Territory 0200, Australia}

\begin{abstract}{
We report on the preparation of entangled two mode squeezed states of yet unseen quality.  Based on a measurement of the covariance matrix we found a violation of the \emph{Reid and Drummond EPR}-criterion  at a value of only $0.36\pm0.03$ compared to the threshold of 1.  Furthermore, quantum state tomography was used to extract a single photon Fock state solely based on homodyne detection, demonstrating the strong quantum features of this pair of laser-beams. The probability for a single photon in this ensemble measurement exceeded 2/3. 
}
\end{abstract}
\maketitle

\paragraph{Introduction.} 
Entanglement plays an ubiquitous role in the field of quantum physics \cite{Horodecki:2009p773}. Light fields are widely used as the carrier for entangled states as they are readily accessible and of high quality in the experiments. Traditionally, quantum opticians are in two complementary camps. One is dealing with discrete variables using individual photons, the other with continuous variables using bright laser beams. Both camps have collected a rich set of tools and techniques in their individual toolboxes. On the one hand, the discrete camp focuses on the preparation and counting of photons. The corresponding theory describing the quantum states and the corresponding measurements is based on finite dimensional Hilbert spaces. Experimental progress has included various sources for single (and multiple) photons\cite{PhysRevLett.69.1516,NeergaardNielsen:2007p912}, sources for entangled pairs and larger groups of entangled photons\cite{PhysRevLett.75.4337,5photEnt}. Numerous  \emph{Bell tests}, quantum teleportation, different aspects of quantum communication were demonstrated\cite{PhysRevLett.49.1804,PhotonTele,BB84}.  On the other hand, the continuous camp focuses on the field aspects of quantum light and uses mainly phase sensitive homodyne detection. In order to accomplish an equivalently simple theory the theorists had to stick to Gaussian fluctuations and operations. This is well justified for the majority of present experiments which have access to moderately strong  nonlinearities.  There are squeezed states, \emph{Einstein-Podolsky-Rosen} (EPR) like entangled states, and an equal variety of quantum communication topics\cite{PhysRevLett.68.3663,Furusawa23101998,PhysRevA.78.012301,DistCont}. A common aim here is to increase the strength and reliability of the entanglement in order to improve the quality of the applications. 

\begin{figure}[ht]
\centering
\includegraphics[width=1\linewidth]{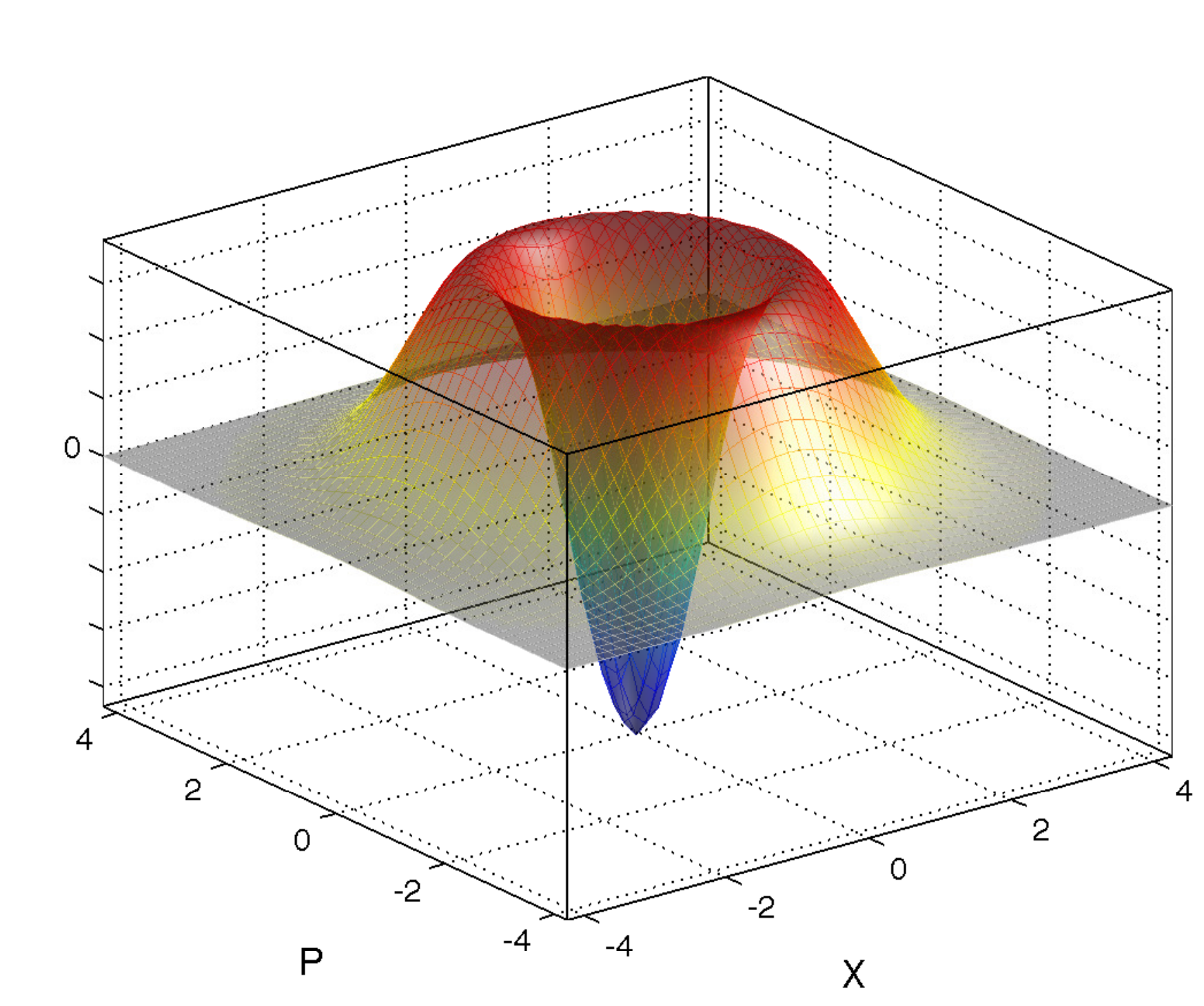}
\caption{Wigner function of a  single photon state reconstructed from a pair of entangled laser-beams. We used quantum state tomography in order to characterise the output state. The reconstruction shown here is based on the inverse radon transform without any correction for detection efficiency or detector dark noise. }
\label{fig:wigner}
\end{figure}

In recent years hybrid experiments have overcome the separation of discrete and continuous variable quantum optics  e.g.~\cite{Ourjoumtsev:2006p883,Lvovsky:2001p853,Ourjoumtsev:2006p879,Wakui:2007p904,NeergaardNielsen:2007p912}. These combine single photon and homodyne detection. The \emph{click}-detector is used to generate trigger pulses displaying a non-Gaussian operation. These trigger pulses are used to gate the homodyne measurement according to the temporal shape of the detected photon, which then produces states of light with impressively strong quantum features. Based on photon pair sources this method was used to generate e.g.~single (few) photon states, single photon added thermal states and single(few) photon subtracted squeezed states\cite{Ourjoumtsev:2006p883,Lvovsky:2001p853,Ourjoumtsev:2006p879,Wakui:2007p904,Parigi28092007,NeergaardNielsen:2007p912}. The latter exhibit a high fidelity with small \emph{Schr\"odinger} cat states. The goal is to show these effects as clearly as possible. However, present single photon detectors suffer either from poor mode (spatial, frequency, temporal) selectivity, limited detection efficiency, noticeable dark noise, lacking photon number discrimination or a combination of these.

The work we present here entirely relies on continuous techniques, i.e.~we use a bright pair of quadrature entangled laser beams and balanced homodyne detection only. A number of technical improvements to our experimental apparatus lead to a previously unseen quality of EPR entangled states as described  by the well  known criterion for bipartite continuous variable entanglement \cite{Einstein:1935p1309,Reid:1988p1372,RevModPhys.81.1727}. For the combined conditional variance we found a value as low as 0.36. Additionally, we demonstrate the proposal of Ralph, Huntington and Symul \cite{PhysRevA.77.063817} reconstructing a single photon Wigner function by post processing our experimental data, which is analogues to the hybrid experiments without actually detecting single photons. This method is similar to the generation of heralded single photons as based on spontaneous parametric downconversion, and was also used for the reconstruction of optical Schr\"odinger Kitten states \cite{kitten}.  However, instead of using a single photon detector on one of the output modes we used homodyne detection in both and exploited the relation $\hat n + \tfrac{1}{2} =\tfrac{1}{4}\left( \hat{X}^2+\hat{P}^2\right)$, with $\hat n $ representing the photon number, $\hat X$ and $\hat P$ for the quadrature operators. This idea was suggested in  \cite{PhysRevLett.85.2035} and applied in  \cite{PhysRevLett.98.153603} for the first time.  Of course, we cannot measure $\hat X$ and $\hat P$ simultaneously, but rather only sequentially. However, relying on our ability to prepare the same state over and over again, in the ensemble average we can use the relation in order to mimic the \emph{real} photon detection. The reconstructed probability for a single photon for the experimental data of Figure \ref{fig:nstat} exceeds 67~\% and can be seen as an overall figure of merit for the efficiency. Note that we did not correct our data for any experimental imperfections, such as quantum efficiencies or dark noise unless indicated otherwise.

\paragraph{Preparation of Entanglement.}

\begin{figure}[ht]
\includegraphics[width=\linewidth]{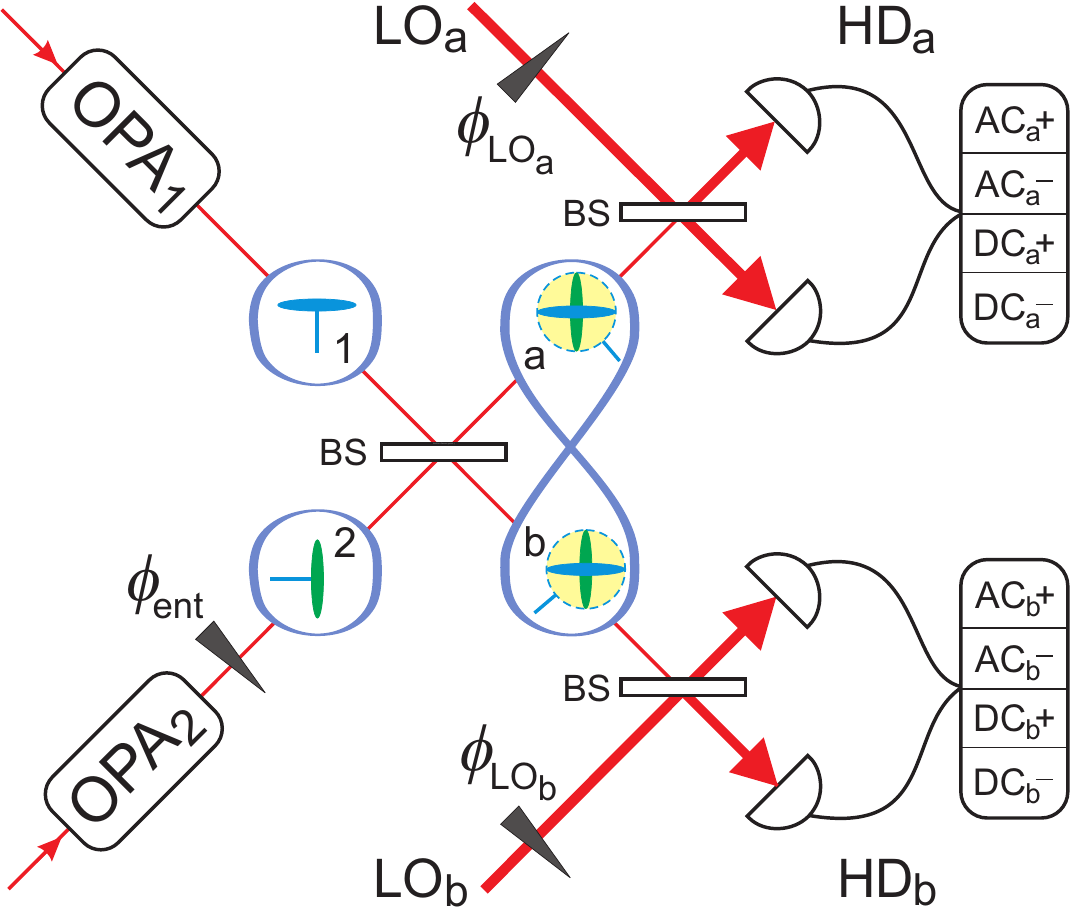}
\caption{An experimental schematics used to generate and analyse a pair of entangled beams.  OPA: optical parametric amplifier, BS: 50:50 beamsplitter, LO: local oscillator, HD: homodyne detection, $\phi$: relative phase between optical fields. }
\label{fig:setup}
\end{figure}
 
A scheme to generate a pair of entangled beams is shown in Fig. \ref{fig:setup}. The main building block is a pair of optical parametric amplifiers (OPA) operating in a de-amplification regime such that a pair of amplitude-squeezed states of light is produced. A dual wavelength laser at 1064 nm and 532 nm was used to drive the experiment. The OPAs are of a bow-tie cavity design with a periodically poled KTP crystal providing the nonlinear interaction. In order to maximise the cavity escape efficiency, faces of the nonlinear crystal were anti-reflection coated using the ion beam sputtering technique providing ultra-low loss for the intra-cavity circulating field. Also the cavity mirrors were custom made with reflectivities exceeding 99.95$\;$\%. The squeezed field is exiting the cavity through an output coupler, which was of 10$\;$\% transmissivity. The two squeezed beams, with almost perfectly equal performance with squeezing levels of approximately -6$\;$dB and an anti-squeezing of 8.5$\;$dB, are mixed on a 50:50 beamsplitter and locked in quadrature; this is keeping their relative phase shift $\phi_{ent}$ of $\pi/2$. This results in the generation of a pair of entangled beams, which are analysed using two balanced homodyne detectors (HD). These detectors had a quantum efficiency of 95$\;$\% and a dark noise clearance of 20$\;\mathrm{dB}$ at the sideband detection frequency of 3$\;\mathrm{MHz}$.  For the single photon reconstruction we used  very weak squeezing of 0.8$\;\mathrm{dB}$ at a sideband frequency of 6$\;\mathrm{MHz}$.

In order to characterise the Gaussian entanglement, we need to
successively measure second order correlations and
anti-correlations between the HD signals, when the HD systems are locked to the phase and amplitude quadratures, respectively.
Measurements of both quadratures are accessible for the two entangled beams depending on the relative phase between
the local oscillator (LO) beam and the analysed field. Hence we locked both HD systems to a half- or full-fringe of the
interference pattern on the HD beamsplitters in order to measure the phase and amplitude quadratures, respectively.

The homodyne detectors' AC and DC electronic signals are recorded by
 a high speed data acquisition system (DAQ) and are used
either to monitor the optical phase of individual beams in real time or for measurements of quantum correlations
 between the entangled fields.  In particular, the phase lock between the two squeezed fields interfering on the entanglement beam splitter (BS) is crucial since it influences the quality of entanglement between the two BS output fields. The error signal for this lock was generated by a combination of HD detectors' DC signals using the DAQ. However, the error signals' DC offset fluctuations, having origin in parasitic interference, were setting the limit for the quality of the entangled fields. We therefore corrected the error signal for these fluctuations using the demodulated HD detectors' AC signals at a specific modulation frequency. The corresponding arithmetics and the control loop itself were implemented in an FPGA. As a result a strong and stable error signal was produced allowing us to access states of high degree of entanglement.

We evaluated several entanglement criteria based on the measured data. For the \emph{Duan inseparability} criterion\cite{Duan:2000p1377,Simon:2000p1472}:
 \begin{equation}\mathcal I=\tfrac{1}{2}\left(  \left\langle \delta \left(\hat X_a + \hat X_b\right)^2 \right\rangle +\left\langle \delta \left(\hat P_a - \hat P_b\right)^2 \right\rangle\right),\end{equation} we found a value of $I=0.3\pm0.02$. We also evaluated the \emph{Reid and Drummond EPR} criterion \cite{Reid:1988p1372,Bowen:2004p1501} based on conditional variances: 
 \begin{equation}
\mathcal E = \Delta^2\hat X_{a|b} \Delta^2\hat P_{a|b}
 \end{equation}
with $\Delta^2\hat X_{a|b}$ denoting the conditional variance. Our data showed $\mathcal E = 0.36\pm0.03$, which displays the lowest value reported so far to our knowledge.

\paragraph{Single photon detection and evaluation scheme.}

The hybrid experiments involving a single photon detector in one of the pair of modes take the homodyne detector data of the other mode into account only when the single photon detector fires. For the reconstruction of the quadrature histograms this means, that most of the homodyne data contributes with a weight of $0$ to the histogram. Only when the single photon detector fires, the weight for the homodyne data is changed to $1$.  
 
\begin{figure}[ht]
\includegraphics[width=\linewidth]{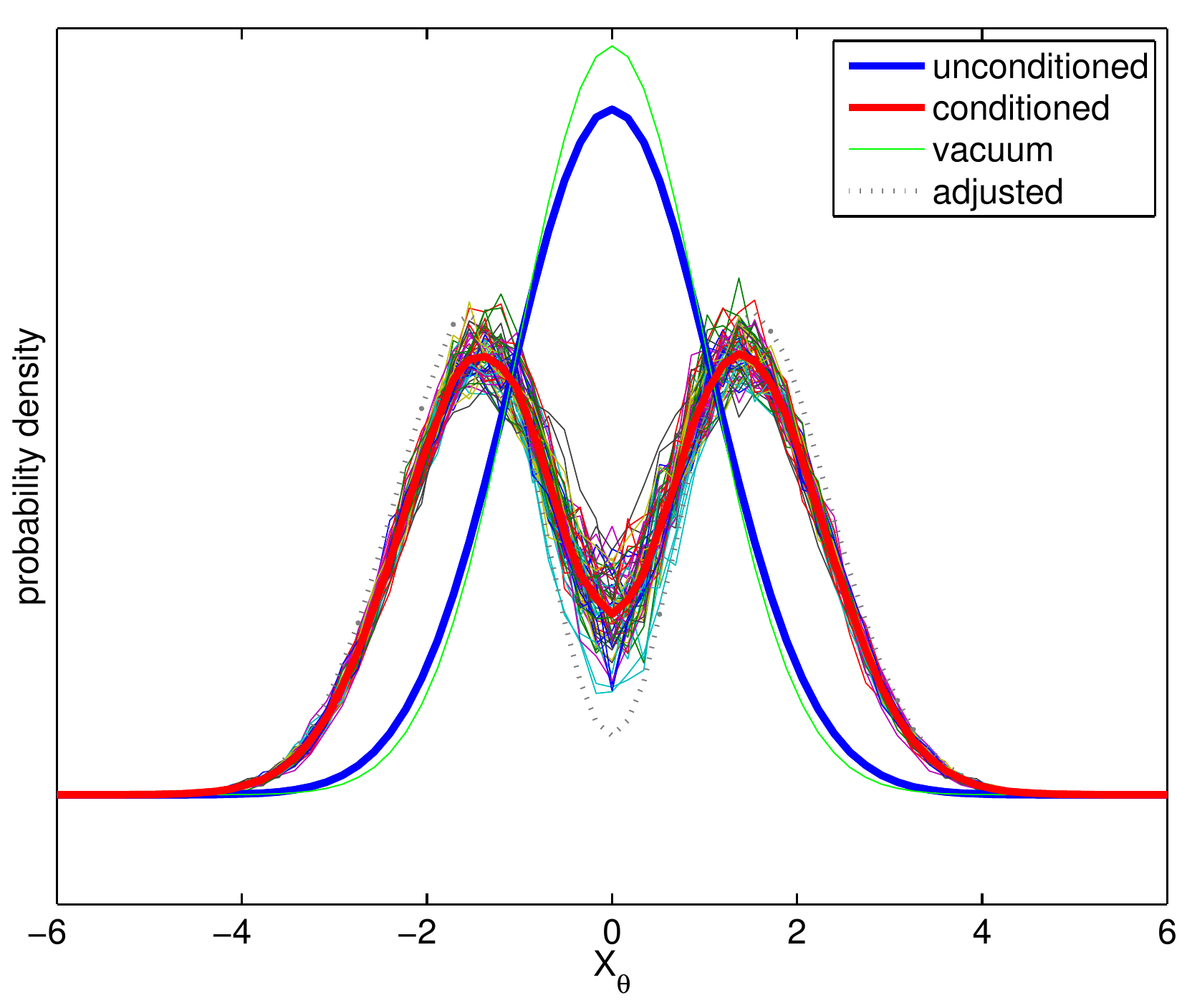}
\caption{Quadrature probability distributions. The thin green curve represents the vacuum state. The thick blue curve shows the average of all quadrature measurements without conditioning. We carefully checked that this is independent of the projection angle. This is also true for the conditioned data, as shown by the thin lines in the background. The noise is due to statistics. The average of these curves is shown as thick red line. The dashed grey line indicates the result of an adjustment for the finite quantum efficiency.}
\label{fig:qhist}
\end{figure}

A pure two mode squeezed state can be written in the Fock basis as: 
\begin{equation} 
\left|\Psi_{TMSS}\right>=\frac{1}{\cosh r}\sum_{n=0}^\infty(-\tanh r)^{n}\left|n\right>_a\left|n\right>_b,
\end{equation}
with $a$ and $b$ for the two modes and $r$ for the squeezing parameter.
Hence, when $n$ photons are found in mode $a$, the other mode $b$ also has $n$ photons. This is how most heralded single photon sources work. There is a \emph{click}-detector in one mode. A click usually occurs, when there was at least one photon hitting the detector. So with every click in one mode there is at least one photon in the other mode.  Now there is a tradeoff for the magnitude of $r$. On one hand the maximum probability of $\tfrac{1}{4}$ for one photon in each mode occurs for $r=\mathrm{arcosh} \sqrt 2$. However, the probability for having two or more photons in this case is just as high, which leads to a highly mixed state. On the other hand for $r\approx 0.1$ the probability for one photon drops to 1$\;$\% but two or more photons now occur one hundred times less likely.

We extend the same idea with a continuous weighting based on data from the homodyne detector, which replaces the single photon detector in our case. 
\begin{eqnarray}
\left\langle \hat n  \right\rangle & = & \tfrac{1}{4}\left\langle \hat X^2 + \hat P^2 -2\right\rangle \nonumber\\
& = & \tfrac{1}{2} \left\langle\left\langle \hat Q^2_\theta -1\right\rangle\right\rangle_\theta, \label{avgqtheta}
 \end{eqnarray}
 with $\langle\rangle_\theta$ denoting the average over all detection phases $\theta$ and $\hat Q_\theta = \cos(\theta)\hat X +\sin(\theta) \hat P$.
 
For the \emph{ordinary} single mode tomography the probability distribution for a given quadrature is reconstructed by taking statistically significant number of measurements $q_a$ of the same quadrature. The range of all possible  measurement results is then divided into a certain number of discrete bins. Each time a measurement result falls into a particular bin, its value is increased by a fixed increment. This means that every single measurement represents the ensemble equally well. In our case the homodyne measurement result $q_b$ of the second mode $b$ of the entangled pair provides some information about the first mode $a$. When building the histogram for mode $a$ instead of using a fixed increment, the bin corresponding to the measurement $q_a$ is increased by the increment $\mathrm di$ which depends on $q_b$: 
\begin{equation}
\mathrm di= q_b^2-1.
\end{equation}

Like for \emph{ordinary} tomography each of the detection phases is measured repeatedly to provide statistical significance. However, during the repeated measurements of mode $a$ the detection quadrature of mode $b$ is varied such that all phases occur uniformly. According to Eqn. \ref{avgqtheta} this means that  {the reconstructed histograms $p_h(q_a)$ give the probability distributions of $p(q_a)$ convoluted by $\langle n_b\rangle_{|q_a}$:
 \begin{equation}
 \label{ph}
p_{h}(q_a)=\langle n_b\rangle_{|q_a} p(q_a)
 \end{equation}
 where $\langle...\rangle_{|q_a}$ is the mean value conditioned to $Q_a=q_a$. We now note that for small squeezing parameters $r$, the probability for having 2 or more photons is small so we can approximate 
  $\langle n\rangle \simeq p(1)$ and hence  $\langle n_b\rangle_{|q_a}\simeq p(1_b|q_a)$.
Injecting this result into (\ref{ph}) and using Bayes law to `invert'  conditional probabilities, we find
 \begin{equation}
 \label{ph2}
p_{h}(q_a)= p(q_a|1_b)\;p(1_b).
 \end{equation}
 Within a renormalisation factor $p(1_b)$, we find the desired probability distribution corresponding ideally to a single photon in $a$ heralded by a `click' in $b$. 
}

\begin{figure}[ht]
\includegraphics[width=\linewidth]{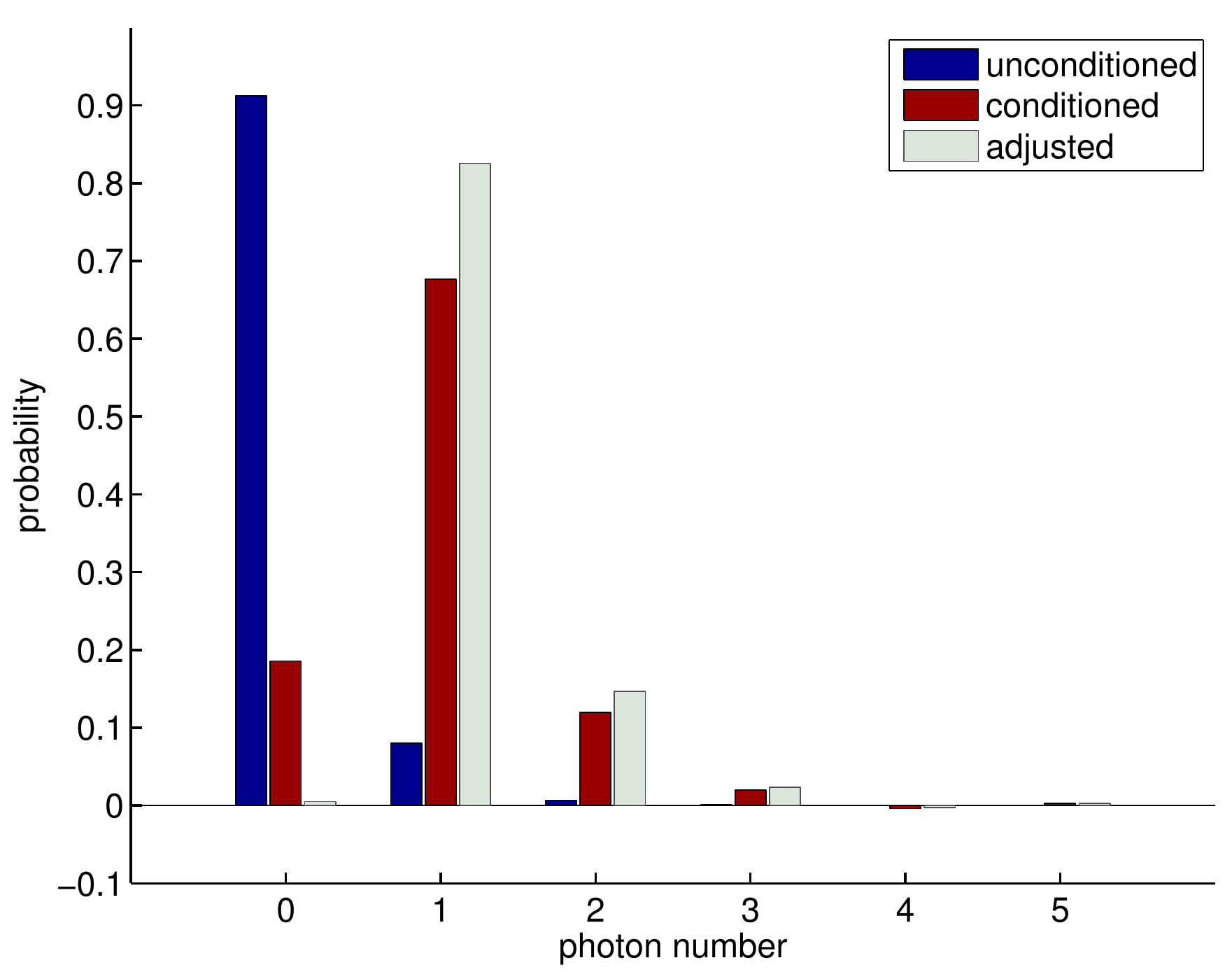}
\caption{Reconstructed photon number distributions. The blue bars are extracted from the measurement data directly. The red bars are the result of the single photon conditioning. The single photon contribution exceeds 67\%, which we would like to consider as a benchmark figure for the overall quality of the experimental apparatus. Note that there is no correction for detection efficiency or dark-noise for these data. The grey bar indicates the result of an adjustment for the finite quantum efficiency. }
\label{fig:nstat}
\end{figure}

The complete characterisation of a general two mode state by means of quantum state tomography requires the measurement of a large number of projection angles and their combinations, which is experimentally challenging. In the particular scenario considered here, the requirements are far less complex. The individual modes of an entangled two mode squeezed state are known to be circular symmetric in quadrature phase space. However, the correlation between the two depends on the relative detection phase.  The conditioning method we use for the single photon projection averages over all detection phases for one of the modes. This implies, that also the dependance of the mutual detection phase is cancelled out. For the reconstruction we ensured, that there actually is no phase dependence in our data. This allows us to base our tomographic reconstruction on a single averaged quadrature distribution, which we assume to be the same for all quadratures. Figure \ref{fig:qhist} shows that this assumption is valid. 

On the experimental side we ensured that for the detection of the conditioning mode $b$ all quadrature projections are represented equally by rapidly scanning the optical phase of the local oscillator with a range adjusted to cover an interval of $2\pi$.  At the same time we slowly varied the detection phase of the detection of mode $a$ over a range much wider than $2\pi$. The data collection time for each histogram was chosen to be short compared to the variation of mode $a$ and also as an integer multiple of the scanning period of mode $b$. That means that each histogram represents a fixed quadrature projection for mode $a$ with a uniform contribution of all projection phases of mode $b$.  
From this procedure we see, that the shape of the reconstructed probability distribution indeed does not depend on the detection phase of mode $a$. Thus, we were able do the state tomography assuming a circular symmetric state without exactly keeping track of the detection phase.  

Based on this probability distribution we used the inverse radon transform technique for the reconstruction of the Wigner function, see Figure \ref{fig:wigner}. Furthermore we used the \emph{pattern function} method of \cite{Leonhardt:1996p1510} in order to reconstruct the Fock base density matrix. The diagonal elements are shown in Figure \ref{fig:nstat}. Clearly, the $n=1$ component is the most dominant one. The residual vacuum component arises from experimental imperfections, such as preparation or detection efficiencies. For a perfect experiment the vacuum component would vanish. We used the capability of the reconstruction method to adjust for the finite efficiency for the grey components of Figure \ref{fig:nstat} and \ref{fig:qhist} . In this sense we consider the residual vacuum component as an overall figure of merit for the quality of the experiment. 

\paragraph{Conclusion.}
The advances in technology and methods enable the implementation of better and more elaborate quantum optics experiments. In our experiment this gives rise to a remarkable progress (see Figure 9 in \cite{RevModPhys.81.1727}) for the violation of the Reid and Drummond EPR criterion as well as the Duan inseparability criterion. Furthermore, sophisticated data acquisition and processing techniques allowed us to apply the proposal of \cite{PhysRevA.77.063817}and  reconstruct a high purity single photon state from our measurement data. This complementary approach to single photon detection could be useful for the implementation of more complex protocols which might be prohibited by the limitations of single photon detection.

\paragraph{Acknowledgement.} 
We would like to acknowledge the financial support by the Australian Research Council through the Centre of Excellence Program (project numbers CE0348178 and CE11E0096) and fellowships and by the European FP7 research project HIDEAS. B.H. appreciates the support by the Alexander von Humboldt-Foundation. We thank Jean-Francois Morizur  for valuable discussions.

\bibliographystyle{bowtitle}
\bibliography{TMS1photV3}

\begin{thebibliography}{}

\bibitem{Horodecki:2009p773}
R.  Horodecki et~al., ``{}\textit{Quantum entanglement}''{}, Rev Mod Phys
  \textbf{81}, 865--942 (2009).

\bibitem{PhysRevLett.69.1516}
T.  Basch\'e et~al., ``{}\textit{Photon antibunching in the fluorescence of a
  single dye molecule trapped in a solid}''{}, Phys. Rev. Lett. \textbf{69},
  1516--1519 (1992).

\bibitem{NeergaardNielsen:2007p912}
J. S.  Neergaard-Nielsen et~al., ``{}\textit{High purity bright single photon
  source}''{}, Optics Express \textbf{15}, 7940--7949 (2007).

\bibitem{PhysRevLett.75.4337}
P. G.  Kwiat et~al., ``{}\textit{New High-Intensity Source of
  Polarization-Entangled Photon Pairs}''{}, Phys. Rev. Lett. \textbf{75},
  4337--4341 (1995).

\bibitem{5photEnt}
Z.  Zhao et~al., ``{}\textit{Experimental demonstration of five-photon
  entanglement and open-destination teleportation}''{}, \textbf{430}, 54--58
  (2004).

\bibitem{PhysRevLett.49.1804}
A.  Aspect, J.  Dalibard and G.  Roger, ``{}\textit{Experimental Test of Bell's
  Inequalities Using Time- Varying Analyzers}''{}, Phys. Rev. Lett.
  \textbf{49}, 1804--1807 (1982).

\bibitem{PhotonTele}
D.  Bouwmeester et~al., ``{}\textit{Experimental quantum teleportation}''{},
  \textbf{390}, 575--579 (1997).

\bibitem{BB84}
C. H.  Bennett and G.  Brassard, ``{}\textit{Quantum Cryptography: Public key
  distribution and coin tossing}''{}, IEEE Conference on Computers, Systems and
  Signal Processing, Bangalore, India 175--179 (1984).

\bibitem{PhysRevLett.68.3663}
Z. Y.  Ou et~al., ``{}\textit{Realization of the Einstein-Podolsky-Rosen
  paradox for continuous variables}''{}, Phys. Rev. Lett. \textbf{68},
  3663--3666 (1992).

\bibitem{Furusawa23101998}
A.  Furusawa et~al., ``{}\textit{Unconditional Quantum Teleportation}''{},
  Science \textbf{282}, 706-709 (1998).

\bibitem{PhysRevA.78.012301}
M.  Yukawa et~al., ``{}\textit{Experimental generation of four-mode
  continuous-variable cluster states}''{}, Phys. Rev. A \textbf{78}, 012301
  (2008).

\bibitem{DistCont}
B.  Hage et~al., ``{}\textit{Preparation of distilled and purified
  continuous-variable entangled states}''{}, \textbf{4}, 915--918 (2008).

\bibitem{Ourjoumtsev:2006p883}
A.  Ourjoumtsev, R.  Tualle-Brouri and P.  Grangier, ``{}\textit{Quantum
  homodyne tomography of a two-photon fock state}''{}, Phys. Rev. Lett.
  \textbf{96}, 213601 (2006).

\bibitem{Lvovsky:2001p853}
A.  Lvovsky et~al., ``{}\textit{Quantum state reconstruction of the
  single-photon Fock state}''{}, Phys. Rev. Lett. \textbf{87}, 050402 (2001).

\bibitem{Ourjoumtsev:2006p879}
A.  Ourjoumtsev et~al., ``{}\textit{Generating optical Schrodinger kittens for
  quantum information processing}''{}, Science \textbf{312}, 83--86 (2006).

\bibitem{Wakui:2007p904}
K.  Wakui et~al., ``{}\textit{Photon subtracted squeezed states generated with
  periodically poled KTiOPO4}''{}, Optics Express \textbf{15}, 3568--3574
  (2007).

\bibitem{Parigi28092007}
V.  Parigi et~al., ``{}\textit{Probing Quantum Commutation Rules by Addition
  and Subtraction of Single Photons to/from a Light Field}''{}, Science
  \textbf{317}, 1890-1893 (2007).

\bibitem{Einstein:1935p1309}
A.  Einstein, B.  Podolsky and N.  Rosen, ``{}\textit{Can Quantum-Mechanical
  Description of Physical Reality Be Considered Complete?}''{}, Physical review
  \textbf{47}, 777 (1935).

\bibitem{Reid:1988p1372}
M. D.  Reid and P. D.  Drummond, ``{}\textit{Quantum correlations of phase in
  nondegenerate parametric oscillation}''{}, Phys. Rev. Lett. \textbf{60}, 2731
  (1988).

\bibitem{RevModPhys.81.1727}
M. D.  Reid et~al., ``{}\textit{Colloquium: The Einstein-Podolsky-Rosen
  paradox: From concepts to applications}''{}, Rev. Mod. Phys. \textbf{81},
  1727--1751 (2009).

\bibitem{PhysRevA.77.063817}
T. C.  Ralph, E. H.  Huntington and T.  Symul, ``{}\textit{Single-photon side
  bands}''{}, Phys. Rev. A \textbf{77}, 063817 (2008).

\bibitem{kitten}
H. M.  Chrzanowski et~al., ``{}\textit{Reconstruction of Optical
  Schr{\"od}inger Kitten States Solely with Continuous Variable Field
  Measurements}''{}, arXiv:1102.5566v1 (2011).

\bibitem{PhysRevLett.85.2035}
T. C.  Ralph, W. J.  Munro and R. E. S.  Polkinghorne, ``{}\textit{Proposal for
  the Measurement of Bell-Type Correlations from Continuous Variables}''{},
  Phys. Rev. Lett. \textbf{85}, 2035--2039 (2000).

\bibitem{PhysRevLett.98.153603}
N. B.  Grosse et~al., ``{}\textit{Measuring Photon Antibunching from Continuous
  Variable Sideband Squeezing}''{}, Phys. Rev. Lett. \textbf{98}, 153603
  (2007).

\bibitem{Duan:2000p1377}
L. M.  Duan et~al., ``{}\textit{Inseparability Criterion for Continuous
  Variable Systems}''{}, Phys. Rev. Lett. \textbf{84}, 2722 (2000).

\bibitem{Simon:2000p1472}
R.  Simon, ``{}\textit{Peres-Horodecki Separability Criterion for Continuous
  Variable Systems}''{}, Phys. Rev. Lett. \textbf{84}, 2726 (2000).

\bibitem{Bowen:2004p1501}
W. P.  Bowen et~al., ``{}\textit{Experimental characterization of
  continuous-variable entanglement}''{}, Phys. Rev. A \textbf{69}, 12304
  (2004).

\bibitem{Leonhardt:1996p1510}
U.  Leonhardt et~al., ``{}\textit{Sampling of photon statistics and density
  matrix using homodyne detection}''{}, Optics Communications \textbf{127}, 144
  (1996).

\end{thebibliography}

\end{document}